\documentclass[11pt]{article}

\newcommand\figref{Fig.~\ref}
\newcommand\tabref{Table~\ref}
\usepackage{multirow}
\usepackage[square,numbers]{natbib}
\usepackage{mathtools}
\usepackage{amsmath,amsfonts,amsthm,bm} 
\usepackage{textcomp}
\usepackage{siunitx}
\usepackage{xcolor}
\usepackage{booktabs}
\usepackage{cases}

\usepackage[switch,columnwise]{lineno}
\usepackage{letltxmacro}
\LetLtxMacro{\originaleqref}{\eqref}
\renewcommand{\eqref}{Eq.~\originaleqref}

\renewcommand{\vec}[1]{\ensuremath{\bm{\mathrm{#1}}}}

\newcommand*\squeezespaces[1]{
  \thickmuskip=\scalemuskip{\thickmuskip}{#1}%
  \medmuskip=\scalemuskip{\medmuskip}{#1}%
  \thinmuskip=\scalemuskip{\thinmuskip}{#1}%
  \nulldelimiterspace=#1\nulldelimiterspace
  \scriptspace=#1\scriptspace
}
\newcommand*\scalemuskip[2]{%
  \muexpr #1*\numexpr\dimexpr#2pt\relax\relax/65536\relax
} 

\usepackage[pdfencoding=auto]{hyperref}
\hypersetup{breaklinks,colorlinks,linkcolor=blue,citecolor=blue}
\usepackage{authblk}
\usepackage{placeins}
\usepackage{layout}
\usepackage{geometry}
\date{}
\begin{document}
\allowdisplaybreaks

\title{PhysRFANet: Physics-Guided Neural Network for Real-Time Prediction of Thermal Effect During Radiofrequency Ablation Treatment}                      

\author[a]{Minwoo Shin}
\author[a]{Minjee Seo}
\author[a]{Seonaeng Cho}
\author[b]{Juil Park}
\author[b]{Joon Ho Kwon}
\author[c,d]{Deukhee Lee}
\author[a]{Kyungho Yoon\thanks{\url{yoonkh@yonsei.ac.kr}}}

\affil[a]{School of Mathematics and Computing, Yonsei University, Seoul, South Korea}
\affil[b]{Department of Radiology, Yonsei University College of Medicine, Seoul, South Korea}
\affil[c]{Center for Healthcare Robotics, Korea Institute of Science and Technology, Seoul, South Korea}
\affil[d]{Yonsei-KIST Convergence Research Institute, Yonsei University, Seoul, South Korea}

\maketitle
\begin{abstract}
Radiofrequency ablation (RFA) is a widely used minimally invasive technique for ablating solid tumors. Achieving precise personalized treatment necessitates feedback information on \textit{in situ} thermal effects induced by the RFA procedure. While computer simulation facilitates the prediction of electrical and thermal phenomena associated with RFA, its practical implementation in clinical settings is hindered by high computational demands. In this paper, we propose a physics-guided neural network model, named PhysRFANet, to enable real-time prediction of thermal effect during RFA treatment. The networks, designed for predicting temperature distribution and the corresponding ablation lesion, were trained using biophysical computational models that integrated electrostatics, bio-heat transfer, and cell necrosis, alongside magnetic resonance (MR) images of breast cancer patients. Validation of the computational model was performed through experiments on \textit{ex vivo} bovine liver tissue. Our model demonstrated a 96\% Dice score in predicting the lesion volume and an RMSE of 0.4854 for temperature distribution when tested with foreseen tumor images. Notably, even with unforeseen images, it achieved a 93\% Dice score for the ablation lesion and an RMSE of 0.6783 for temperature distribution. All networks were capable of inferring results within 10 ms. The presented technique, applied to optimize the placement of the electrode for a specific target region, holds significant promise in enhancing the safety and efficacy of RFA treatments.
\end{abstract}

\makeatletter\def\Hy@Warning#1{}\makeatother
\maketitle

\section{Introduction}\label{sec:1}

Radiofrequency ablation (RFA) stands as a pivotal technique in the field of interventional medicine due to its ability to treat various medical conditions, particularly tumors, with minimal invasiveness \cite{McDermott2013-tt,CCO33396,00767571-201837030-00002,yap2021}. RFA uses the heat generated from electrical currents delivered through percutaneously inserted electrodes to induce coagulative necrosis in pathological tissues \cite{Peek2017-pi}. Its significance lies in its effectiveness in eradicating a tumor while preserving healthy surrounding tissues, thus offering advantages compared to the traditional surgical resection, such as quick recovery, high efficacy, low complication rate, and cost-effectiveness \cite{Lim2000-lp,Zhu2013,Berjano2006}.

To achieve precise treatment, interventional radiologists adjust clinical settings such as applied power, current duration, and the insertion trajectory of electrodes \cite{Lee2018Review,Widmann2009-dx,MULIER2020145}. However, quantitatively predicting the extent of ablation remains a challenging task, as treatment outcomes are influenced not only by device settings but also by individual-specific characteristics and the surgeon's expertise. In general planning procedures, ablation shapes are commonly assumed to be ellipsoids, based on manufacturer-provided ablation size guidelines designed primarily for homogeneous tissue. This approach overlooks patient-specific factors crucial for achieving personalized and precise RFA treatment \cite{Jiang2010,McCreedy2006-tl}.

Theoretical models and computer simulations serve as crucial tools in predicting the ablated region and improving therapeutic outcomes for individual patients by providing vital information into the electrical and thermal behaviors during RFA procedure \cite{Berjano2006,bertaccini2007,barauskas2008,shao2017,OOI201912}. The outcomes of temperature distribution and the corresponding ablation lesion are obtained through analysis of multi-biophysics models, which represent electrical conduction, bio-heat transfer, and cell necrosis process \cite{kroger2006,shahidi1994,singh2018}. However, achieving accurate predictions demands a significant computational cost, thereby imposing constraints on its practical application in clinical settings \cite{Voglreiter2018}.

Several efforts have been made to enhance simulation speed for the intra-operative utilization of computational methods during RFA procedures \cite{Mariappan2017,kath2019,Hoffer2022}. These works have achieved significant reductions in computation time owing to the use of massively parallel computing on graphics processing units (GPU). However, even with these advancements, the simulation times still range from a few seconds to minutes, depending on the size of the simulation domain. Furthermore, prolonged simulation times become necessary when utilizing more sophisticated models, such as coupled multi-physics analysis \cite{akbari2021} and considering the phase change effect of biological tissue \cite{abraham2007}.

Recently, there has been an emerging trend to leverage deep learning algorithms for the efficient representation of complex physical phenomena that pose challenges for characterization through traditional mathematical analysis \cite{lutter2019,mendizabal2020,choi2022,salehi2022,shin2023}. Within the domain of RFA treatment, a noteworthy study by \cite{8886409} exemplifies the potential of implementing real-time estimation of RFA lesion depth using a machine learning model combined with a statistical merging approach. The proposed system utilized multi-frequency impedance measurement data, achieving accuracy at the millimeter resolution. However, it is limited by providing only one-dimensional depth information of RFA lesions in a simplistic tissue representation model. For the practical efficacy of such deep learning models in clinical settings, it is essential to accurately represent the multi-physics phenomena arising from the intricate morphologies of lesions.

In this paper, we present PhysRFANet, a set of deep neural networks designed to learn from computational simulations that capture the multi-physics phenomena inherent in RFA procedures. The training data were generated through computational simulations of electrostatic, bio-heat transfer, and cell necrosis models using magnetic resonance (MR) images obtained from 11 actual breast cancer patients. The accuracy of the computational model was validated through experiments using \textit{ex vivo} bovine liver tissue. We developed a total of six network models based on three different architectures, each serving two specific purposes: predicting temperature distribution and the corresponding ablation lesion. The performance of these network models was evaluated using various metrics. Additionally, to evaluate the robustness and generality of our model, we conducted further assessments using two new sets of patient data that were entirely excluded from the training process. The significance of the proposed model is summarized as follows:
\begin{itemize}
    \item The computational biophysical model reflects the geometric characteristics of an individual patient's tumor through a magnetic resonance (MR) image.
    \item Prediction accuracy of the computational model was validated using \textit{ex vivo} bovine liver tissue.
    \item Neural network models were trained on the results of the computational model.
    \item Real-time personalized predictions of thermal distribution and cell necrosis regions are available, given the tumor geometry and electrode placement.
\end{itemize}

\section{Computational model}\label{sec:2}
In this section, we address the formulation of our computational model, which includes a geometric representation of tumor tissue, an electrical field calculation, a bio-heat transfer algorithm, and a cell necrosis model.

\subsection{Tumor geometric model via MR image}\label{sec:2.1}
\begin{figure}
    \centering
    \includegraphics[width=0.8\textwidth]{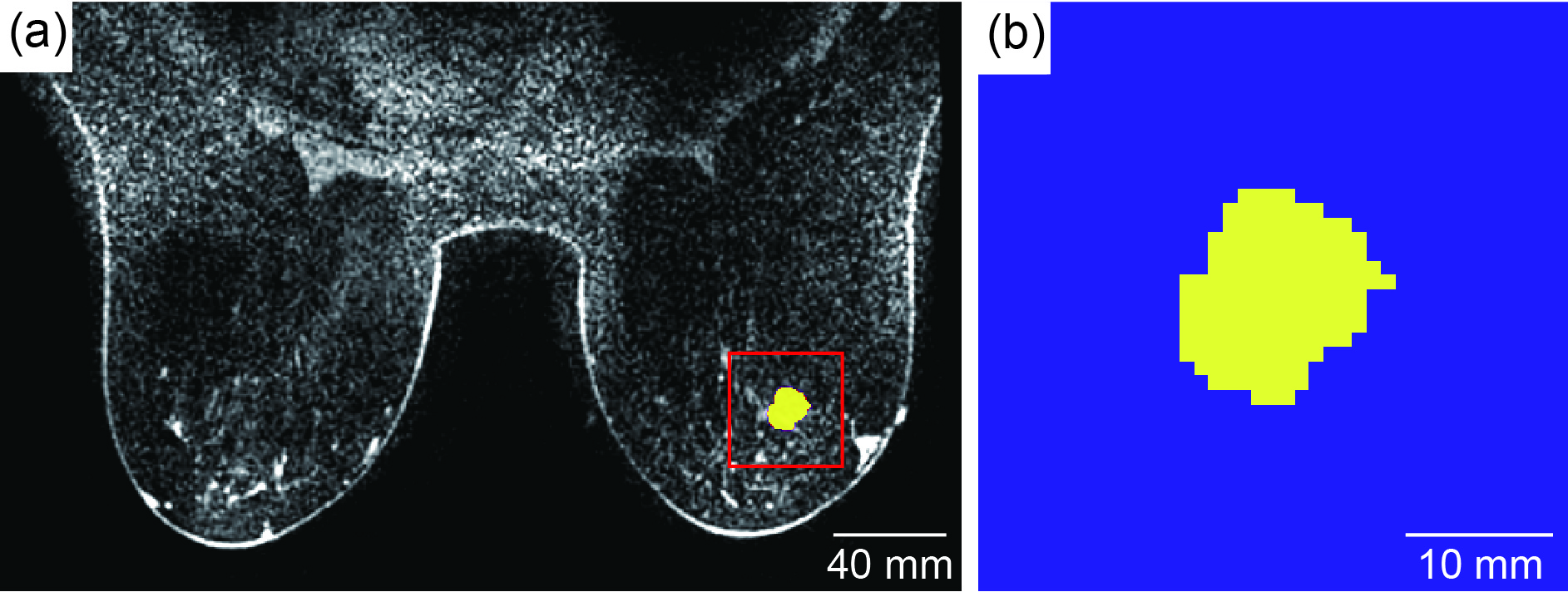}
    \caption{An example of tumor geometric modeling from an MR image. (a) Overlay of tumor segmentation result (yellow region) onto the corresponding axial view MR image. The red box shows the boundary of the cropped field-of-view, and (b) shows the segmentation image in the field-of-view. The yellow region represents the tumor area, while the blue region is normal tissue.}
    \label{fig:MRimage}
\end{figure}

MR images of thirteen breast cancer patients from a publicly available dataset \cite{saha2018} were utilized to model tumor geometry. The datasets consisted of dynamic contrast enhanced (DCE) MR images and the corresponding segmentation images which were annotated by radiologists to depict the cancerous regions for each patient. All images were resampled to isotropic voxels of $1.0\times1.0\times1.0$ mm$^3$ using the Lanczos filter algorithm \cite{burger2009}. For computational efficiency, the simulation domain for each patient was defined by cropping it to the field-of-view centered at the volumetric center of the tumor region with a size of $40\times40\times40$ mm$^3$. The field-of-view size was sufficient to encompass the entire volume of the segmented tumor ($3,021\pm2,789$ mm$^3$, N=13). A visual representation of the tumor geometric model using MR images is provided in \figref{fig:MRimage}.

\subsection{Electro-static analysis for electrical field}\label{sec:2.2}
The quasi-static version of Maxwell’s equation is used to compute the resistive heating during RFA. This is relevant because, within the RFA frequency range of 450 to 550 kHz, the electric field's wavelength is substantially larger than the size of the active electrode, as noted in \cite{Singh2017}. Thus, the electric potential is determined by solving the generalized Laplace equation
\begin{linenomath}
\begin{equation}
    \label{eq:1}
    \nabla\cdot\left(\sigma\nabla V\right)=0 \ \text{in} \ \Omega,
\end{equation}
\end{linenomath}
where $\sigma$ is the electrical conductivity (S/m), $V$ is the electric potential (V) with the following boundary conditions, and $\Omega$ is a volumetric domain for analysis \cite{Schumann2011}: 
\begin{linenomath}
\begin{subequations}
\label{eq:2}
\begin{numcases}{}
   V(\vec{x})=V_\text{p}, & $\vec{x} \in \Gamma_\text{e}$, \label{eq:2a}\\
   \vec{n}(\vec{x})\cdot\nabla V(\vec{x})=0, & $\vec{x} \in \Gamma_\text{s}$
\end{numcases} 
\end{subequations}
\end{linenomath}
where $V_\text{p}$ is the applied power (W), $\Gamma_\text{e}$ is the boundary of the domain covered by the electrodes, and $\Gamma_\text{s}$ is the exterior surface boundary of simulation space.

The finite element formulation of \eqref{eq:1} and (\ref{eq:2}) is as follows:
\begin{linenomath}
\begin{equation}
\label{eq:3}
\begin{split}
    &\int_{\Omega}{\sigma\delta\vec{V}^T\left(
    \frac{\partial\vec{N}^T}{\partial{x}}\frac{\partial\vec{N}}{\partial{x}}+\frac{\partial\vec{N}^T}{\partial{y}}\frac{\partial\vec{N}}{\partial{y}}+\frac{\partial\vec{N}^T}{\partial{z}}\frac{\partial\vec{N}}{\partial{z}}
    \right)\vec{V}}\,d\Omega=0,
\end{split}
\end{equation}
\end{linenomath}
where $\vec{V}$ is the discretized electrical potential at finite element nodes, $\vec{N}$ is a linear interpolation matrix (i.e., $V=\vec{N}\vec{V})$, and $\delta$ indicates a variational symbol. Through the standard finite element solution procedure with \eqref{eq:3}, the solution of electrical potential is obtained.

The volumetric heat source due to resistive heating during RFA, denoted by $Q_\text{r}$ (W/m$^3$), is quantified by
\begin{linenomath}
\begin{equation}
    \label{eq:4}
    Q_\text{r}=\sigma|\nabla V|^2.
\end{equation}
\end{linenomath}
The calculated $Q_\text{r}$ is utilized as the heat input for the bio-heat transfer analysis.

\begin{figure}[t!]
    \centering
    \includegraphics[width=0.29\textwidth]{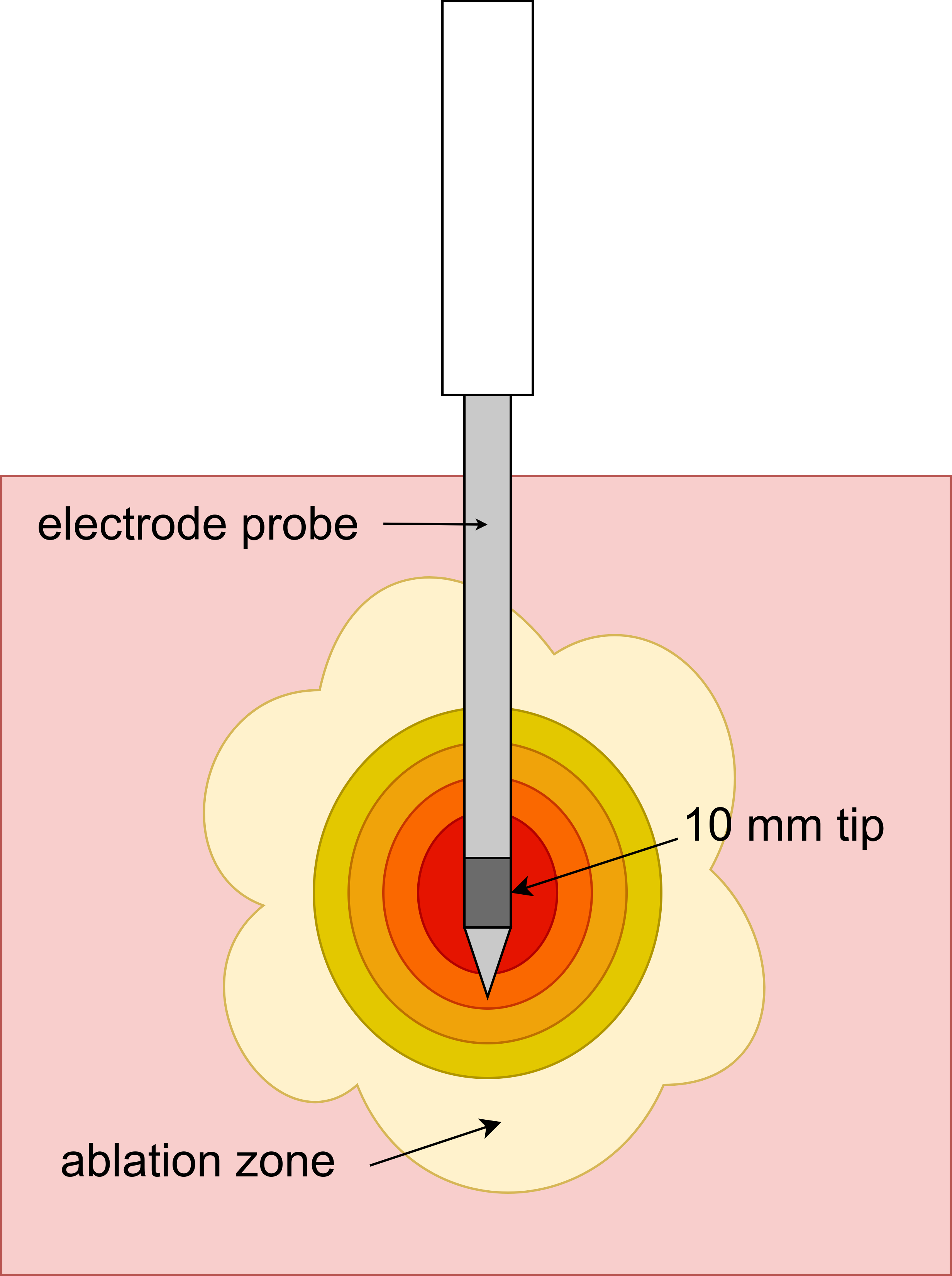}
    \caption{The illustration represents a single-needle ablation electrode utilized for tumor ablation. A source voltage applied to the conductive tip facilitates the therapeutic intervention.}
    \label{fig:RFA}
\end{figure}

Finite element models were constructed based on the obtained tumor geometry in Section \ref{sec:2.1}. The meshes corresponding to the tumor tissue were assigned an electrical conductivity of 4 S/m, while the finite elements classified as normal tissue had an electrical conductivity of 0.4 S/m \cite{zhao2013}. The input voltage generated by an electrode tip with a length of 10 mm and a diameter of 1 mm was modeled by applying boundary conditions in \eqref{eq:2a} to all nodes encompassed by the electrode. Located at the end of the needle, the electrode tip releases high-frequency alternating current into the surrounding tissue, generating heat through radiofrequency energy. An illustration of a single-needle ablation electrode is presented in \figref{fig:RFA}.

\subsection{Bio-heat transfer analysis for heat propagation}\label{sec:2.3}
Pennes bioheat model is used to model the bio-heat transfer inside the breast tissue during RFA \cite{Pennes1948}:
\begin{linenomath}
\begin{equation}
\label{eq:5}
    \rho c \frac{\partial T}{\partial t}=\nabla\cdot(k\nabla T)+\rho_\text{b} c_\text{b} \omega_\text{b} (T_\text{b}-T)+Q_\text{m}+Q_\text{r},
\end{equation}
\end{linenomath}
where $\rho$ is the density of tissue (kg/m$^3$), $c$ is the specific heat capacity of tissue (J/kg/K), $T$ is temperature (\SI{}{\degreeCelsius}), $t$ is ablation time, $k$ is thermal conductivity of tissue (W/m/K), $\rho_\text{b}$ is the density of blood, $c_\text{b}$ is the specific heat of blood, $\omega_\text{b}$ is the perfusion rate of blood, $T_\text{b}$ is the temperature of blood (\SI{37}{\degreeCelsius}), $Q_\text{m}$ is metabolic heat generation (W/m$^3$), and $Q_\text{r}$ is resistive heat generation obtained from \eqref{eq:4} (W/m$^3$).
 
To numerically solve \eqref{eq:5}, the finite difference time domain (FDTD) scheme is used \cite{Berjano2006,Labonte1994,kim2023laser},
\begin{linenomath}
\begin{equation}\squeezespaces{0.01}
\begin{split}
    &\frac{\rho_{i,j,k} c_{i,j,k}}{\Delta t}\left(T^{t+\Delta t}_{i,j,k}-T^{t}_{i,j,k}\right)\\
    &=\frac{k_{i,j,k}}{\Delta s^2}\left[\left(T^{t}_{i+1,j,k}-T^{t}_{i,j,k}\right)-\left(T^{t}_{i,j,k}-T^{t}_{i-1,j,k}\right)\right]\\
    &+\frac{k_{i,j,k}}{\Delta s^2}\left[\left(T^{t}_{i,j+1,k}-T^{t}_{i,j,k}\right)-\left(T^{t}_{i,j,k}-T^{t}_{i,j-1,k}\right)\right]\\
    &+\frac{k_{i,j,k}}{\Delta s^2}\left[\left(T^{t}_{i,j,k+1}-T^{t}_{i,j,k}\right)-\left(T^{t}_{i,j,k}-T^{t}_{i,j,k-1}\right)\right]\\
    &+\rho_\text{b} c_\text{b} \omega_\text{b} \left(T_\text{b}-T^{t}_{i,j,k}\right)+Q_{\text{m};i,j,k}+Q_{\text{r};i,j,k},
\end{split}
\end{equation}
\end{linenomath}
where $\Delta t$ is the discretized time interval, $\rho_{i,j,k}$, $c_{i,j,k}$, $T_{i,j,k}$, $k_{i,j,k}$, $Q_{\text{m};i,j,k}$, and $Q_{\text{r};i,j,k}$ are respectively the density, specific heat capacity, temperature, thermal conductivity, metabolic heat generation, and resistive heat generation of tissue at the grid index $i,j,k$ of the simulation domain, and $\Delta s$ is the spatial discretization interval (i.e., grid interval).

Based on the segmented MR image in Section \ref{sec:2.1}, thermal properties of the tumor tissue (density of 1,050 kg/m$^3$, specific heat capacity of 3,770 J/kg/K, thermal conductivity of 0.48 W/m/K), normal tissue(density of 911 kg/m$^3$, specific heat capacity of 2,348 J/kg/K, thermal conductivity of 0.21 W/m/K), blood perfusion (blood density of 1,050 kg/m$^3$, blood specific heat capacity of 3,617 J/kg/K, blood perfusion rate of 5.3 s$^{-1}$ for tumor tissue, blood perfusion rate of 0.2 s$^{-1}$ for normal tissue), and metabolic heat (400 W/m$^3$ for normal tissue and 13,600 W/m$^3$ for tumor tissue) were assigned to the simulation \cite{singh2018}. The initial temperature of the tissue and blood was set to 37$^\circ C$. The simulations were conducted by applying the resistive heat (i.e., $Q_{\text{m};i,j,k}$) for 180 seconds with a time resolution (i.e., $\Delta t$) of 0.1 seconds and calculating the temperature distribution of the entire spatial domain at each time interval.

\subsection{Cell necrosis model}\label{sec:2.4}
The thermal damage of biological tissue ($\Psi$) is quantified using the first-order Arrhenius rate equation as follows \cite{singh2018}:
\begin{linenomath}
\begin{equation}
    \Psi(t)=\int_{0}^{t}{A\exp\left({\frac{-E_\text{a}}{RT(t)}}\right)}\,dt,
\end{equation}
\end{linenomath}
where $A$ is the frequency factor ($=1.18\times10^{44}$), $E_\text{a}$ is the activation energy for irreversible damage reaction ($=3.02\times10^5$ J/mol), $R$ is the universal gas constant ($=8.3134$ J/mol/K), $T$ is the temperature (K) of the corresponding tissue. 

The temperature distribution for each time step obtained in Section \ref{sec:2.3} was utilized to evaluate the cumulative tissue damage values of $\Psi(t)$ for every voxel. To delineate irreversible thermal damage, we applied a threshold of $\Psi(t)>1$, classifying values exceeding this threshold as the damaged tissue region and values below it as the viable tissue region \cite{henriques1947}.

\subsection{Experimental validation}\label{sec:2.5}
\begin{figure*}[htpb!]
    \centering
    \includegraphics[width=0.72\textwidth]{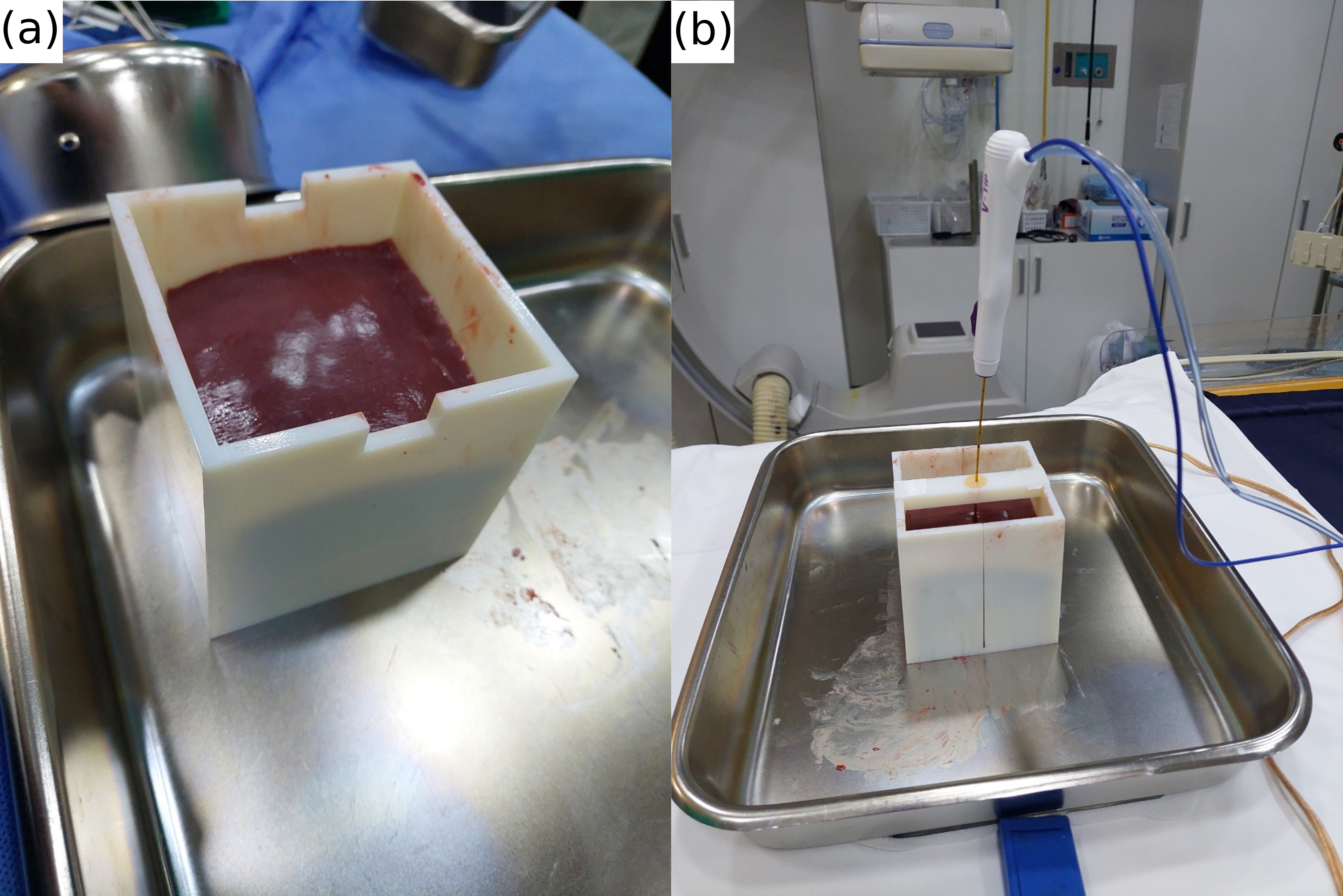}
    \caption{Experimental setup for \textit{ex vivo} tissue RFA. (a) A bovine liver sample is cut into a cuboidal shape and carefully placed within a 3D-printed container, (b) enabling accurate ablation at its center.}
    \label{fig:experiment_setup}
\end{figure*}

In this section, we describe the experiment setup for RFA, designed to validate the accuracy and reliability of our RFA computational model. Freshly excised bovine liver, procured on the day of the experiment from a local butcher shop, served as the experimental tissue. Ten cuboid-shaped samples, each measuring approximately $10\times10\times10$ cm$^3$, were prepared by cutting the bovine liver to fit within a custom-designed 3D-printed container shown in \figref{fig:experiment_setup}(a). This container is designed to facilitate precise RF ablation at the center of the liver sample, followed by a subsequent halving of the sample at the same location, as illustrated in \figref{fig:experiment_setup}(b). 

A 17-gauge monopolar electrode with a tip length of 10 mm (V-tip, RF medical, Seoul, Korea) was inserted to a depth of approximately 5 cm, positioned at the center of the liver sample. Radiofrequency waves were emitted from the electrode tip into the surrounding tissue for 3 minutes, utilizing a 200 W generator (M-3004, RF medical, Seoul, Korea). After completing the ablation, the electrode was removed, and the bovine liver sample was bisected to obtain cross-sectional plane images. Distinct areas of coagulative necrosis were revealed in these images, identified and segmented using the ``Segment Anything in Medical Images'' code \cite{ma2023segment,SAM} for subsequent quantitative analysis.

Through the obtained experimental results, the proposed computational model was validated. A homogeneous tissue space with a resolution of 1 mm, spanning $40\times40\times40$ mm$^3$, was utilized, with an electrode placed at the central location of the simulation space. The electrical conductivity of the bovine liver tissue was set to 0.69 S/m \cite{fuentes2010}. The thermal properties of the tissue included a density of 1079 kg/m$^3$, a specific heat capacity of 3415.0 J/kg/K, and a thermal conductivity of 0.5 W/m/K \cite{silva2020}; blood perfusion and metabolic heat terms were disregarded. The heat generated by radiofrequency energy was applied over a total duration of 180 seconds, using a time step of 0.1 seconds, starting from an initial temperature of 20 $^\circ C$.

\section{Neural network models}\label{sec:3}
In this section, we employed three network models with different architecture for performance comparisons. Specifically, for the purpose of basic performance benchmarking, the encoder-decoder based CNN (EDCNN) architecture was employed. The U-Net architecture, with its symmetric encoder-decoder structure complemented by skip connections, is our primary focus in this research. This design is known for its use of skip connections that combine deep coarse-grained feature maps from the decoder and the shallow fine-grained feature maps from the encoder, effectively enhancing target details \cite{Drozdzal2016,He2016,Huang2017,Hariharan2015,Lin2017feature,zhou2018unet}.

\subsection{Training and test data generation}\label{sec:3.1}
The datasets used to train and test the network model are derived from the computational model described in Section \ref{sec:2}. To ensure a diverse dataset, we conducted RFA simulations for each breast tumor MR image, utilizing 500 randomly placed electrode tip locations and directions within the tumor regions. The segmented breast tumor images (size of $41\times41\times41$) along with binary images (size of $41\times41\times41$) representing the placement of electrode tip were utilized as the network inputs, and either the corresponding coagulative tissue necrosis zones (size of $41\times41\times41$) or temperature distributions (size of $41\times41\times41$) obtained from RFA simulation were respectively employed for the network output.

For the training datasets, we collected 5,500 RFA simulation results derived from eleven distinct breast tumor images (BT1-BT11) from different patients. This data was then randomly split into 5,000 samples for training and 500 for testing purposes. Out of the 5,000 training data samples, 200 were allocated to the validation set for fine-tuning. On the other hand, to evaluate the network's performance on unseen MR images of breast tumors from a practical perspective, we utilized 500 RFA simulation results sourced from two distinct tumors (BT12 and BT13), each from a separate patient, as another test dataset. For a better understanding, please refer to \tabref{tab:data_split}.

In the training of our PhysRFANet, we carefully selected various hyperparameters to optimize performance. In configuring our model's optimization strategy, we selected the Adam optimizer, initialized with a learning rate of 0.001, and set the batch size to 16 for efficient processing. To fine-tune the learning rate during training, we implemented the ``ReduceOnPlateau'' learning rate scheduler. This scheduler was configured to reduce the learning rate by a factor of 0.5 if no improvement is observed in the valid loss over 5 epochs. This approach allows for dynamic adjustment of the learning rate based on the model's performance. The training was conducted over 100 epochs. All training was performed on a single GPU, utilizing its parallel processing capabilities to expedite the training process.

\begin{table}[!b]
    \caption{Splitting of foreseen and unforeseen test datasets.}
    \centering
    \begin{tabular}{lcc}
    \toprule
          & training dataset & test dataset\\
          \midrule
         foreseen  & \multirow{2}{*}{5000 (BT1-BT11)}& 500 (BT1-BT11)~~~\\
         unforeseen&                                 & 500 (BT12-BT13)\\
         \bottomrule
    \end{tabular}
    \label{tab:data_split}
\end{table}

\subsection{Network architectures for the coagulative necrosis}\label{sec:3.2}
In this section, we provide a thorough overview of the PhysRFANet architectures, showcasing them with visual diagrams complemented by in-depth explanations. Visual explanation is provided in \figref{fig:dmg_networks}. Each component and layer will be detailed, emphasizing their specific roles and contributions to the overall system, as evidenced by ablation studies.

The EDCNN model, shown in \figref{fig:dmg_networks}(a), is a CNN specifically tailored for 3D data. It consists of an encoder and a decoder: the encoder is a series of 3D convolutional layers, each followed by batch normalization and ReLU activation, designed to extract and compress features from the input. The encoder uses strides to reduce spatial dimensions, akin to downsampling. The decoder part, aimed at reconstructing the data to a higher resolution, employs transposed 3D convolutional layers, also accompanied by batch normalization and ReLU activations. In its forward pass, the model concatenates two types of input data, indicating its capability to handle dual input modalities, a common scenario in medical imaging tasks. The final layers reduce the channel dimension to one to match the size of the input. Additionally, the model adopts specific weight initialization strategies for its layers, indicating a focus on efficient training convergence.

The U-Net architecture, shown in \figref{fig:dmg_networks}(b), includes the decoder, mid-layers, and encoder classes. The decoder performs downsampling, mid-layers process features at the network bottleneck, and the encoder handles the upsampling part. These modules use convolutional layers, batch normalization, and activation functions suitable for processing 3D data. The U-Net architecture concatenates and processes inputs through the downscaling, middle, and upscaling modules. The architecture is designed for tasks that involve volumetric data.

The Attention U-Net model, shown in \figref{fig:dmg_networks}(c), is a sophisticated neural network architecture that blends the U-Net structure with self-attention mechanisms. This design is especially suitable for complex applications like medical image analysis, which require both local and global contextual understanding. The network features self-attention modules that implement multi-head attention for capturing long-range dependencies within the data, accompanied by layer normalization and feedforward networks for further refinement. The architecture is structured along a U-Net framework, with the encoder constituting the downsampling path. These modules employ 3D convolutions, batch normalization, and LeakyReLU activations to incrementally extract complex features while reducing spatial dimensions. In the middle of the network, mid-layers process these features, employing skip connections for added efficiency. The upsampling path, composed of a decoder, increases the spatial dimensions and integrates features from the downsampling path, enhancing detail retention. Crucially, self-attention modules are interspersed in the upsampling path, augmenting the network's focus on salient features. The forward method defines the workflow, starting with the concatenation of two input data types and proceeding through the successive network layers, with optional interpolation of input and output to specific dimensions.

\subsection{Network architectures for the temperature distribution}\label{sec:3.3}

\begin{figure*}[htpb!]
    \centering
    \includegraphics[width=\textwidth]{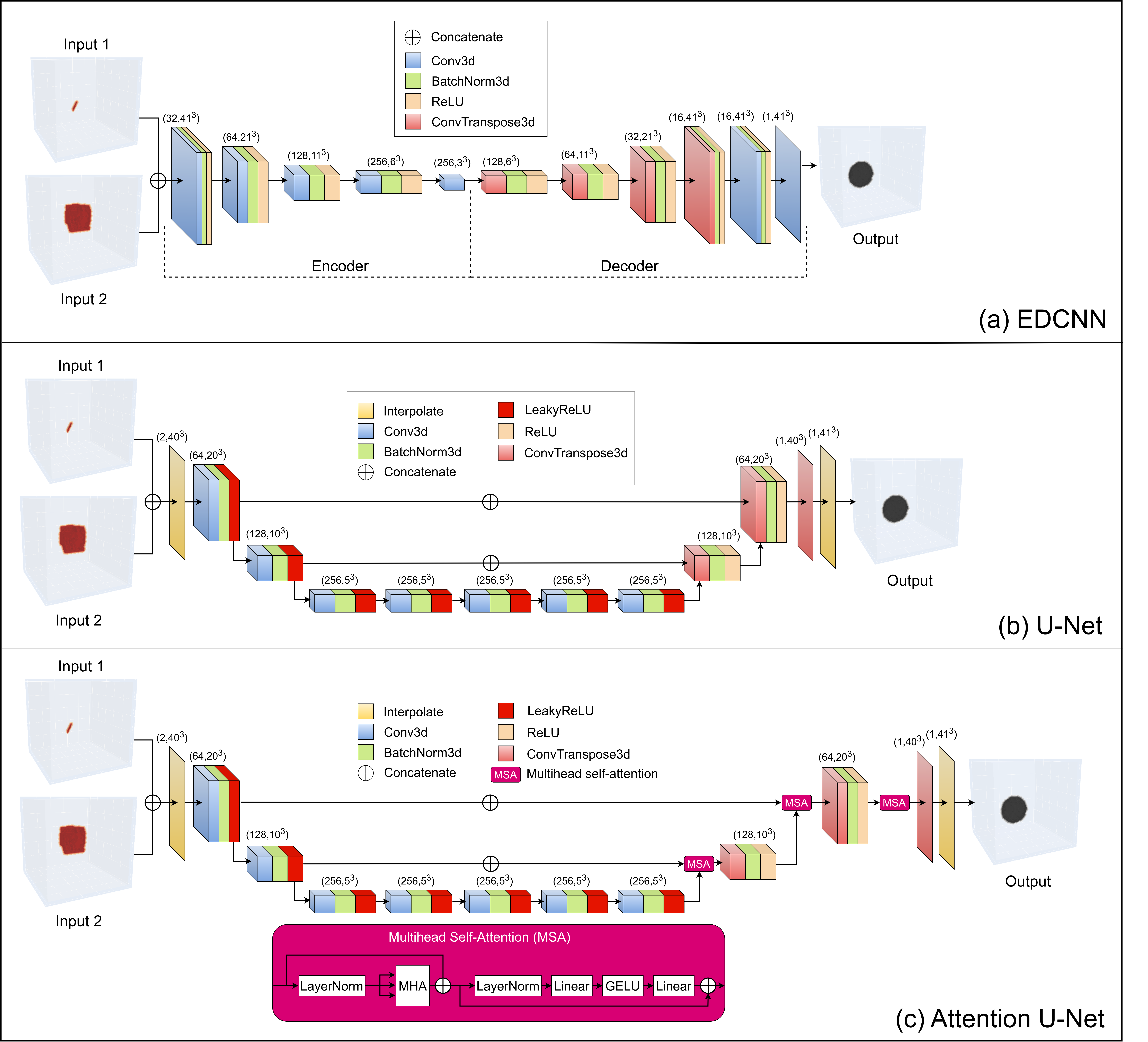}
    \caption{Three network architectures: (a) EDCNN, (b) U-Net, and (c) multihead-Attention U-Net architectures. The input 1 is the geometry of the electrode tip and the input 2 is the segmented breast tumor from the MR image. The output is the predicted ablation lesion zone. The Dice loss function is used for all network training.}
    \label{fig:dmg_networks}
\end{figure*}

\begin{figure*}[htpb!]
    \centering
    \includegraphics[width=\textwidth]{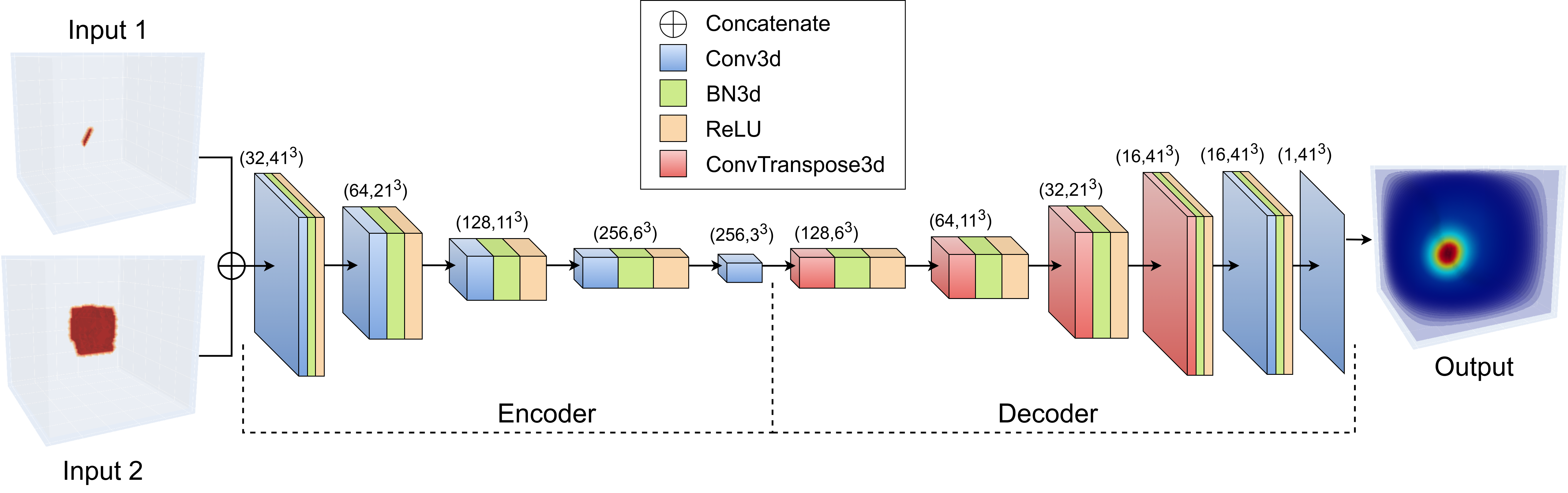}
    \caption{EDCNN architecture, where Input 1 comprises the electrode tip geometry and Input 2 is the segmented breast tumor from the MR image. The output is the predicted temperature distribution map. The U-Net and Attention U-Net architectures employed for temperature distribution prediction are identical to those in \figref{fig:dmg_networks}(b) and (c), with the primary difference being the output, which is now the temperature distribution. The combined loss function \eqref{eq:loss} with $(\alpha,\beta,\gamma)=(0.8,0.1,0.1)$ is used for all network training for the temperature distribution.}
    \label{fig:networks1}
\end{figure*}

The same network architectures described in Section \ref{sec:3.2} are utilized for analyzing the temperature distribution. The primary distinction between the networks shown in \figref{fig:dmg_networks}(a) and \figref{fig:networks1} lies in their outputs. Otherwise, the networks are largely similar, with the main differences being in their respective outputs. As in the previous section, ablation studies will be conducted to determine how the network's performance is impacted by the change of various components.

\subsection{Loss function}\label{sec:3.4}
For the network training of the ablated lesion zone, we employ the Dice loss function, which is defined as follows:
\begin{linenomath}\begin{equation}
    \text{Dice}(x,\widetilde{x}) = 1-\frac{2\sum_{i=1}^{n}{x_{i}\widetilde{x}_{i}}}{\sum_{i=1}^{n}{x_{i}^2}+\sum_{i=1}^{n}{\widetilde{x}_{i}^2}},
\end{equation}\end{linenomath}
where $\widetilde{x}_i$ and $x_i$ represent the components at the $i$-th position in the ground truth and the corresponding ablation lesion prediction, respectively.

For the network training of the temperature distribution, we utilize a loss function that combines mean squared error (MSE), weighted MSE, and the Dice$_{(>50)}$ loss. To introduce a weighted MSE loss function, define a mask $m$ as follows:
\begin{linenomath}
\begin{equation}
    m_i=\begin{cases}
        1, \quad\text{if }x_i>50^\circ C,\\
        0, \quad\text{otherwise}.
    \end{cases}
\end{equation}
\end{linenomath}
where $\widetilde{x}_i$ and $x_i$ denote the elements at the $i$-th position in the ground truth and the corresponding temperature distribution prediction, respectively. We implemented the weighted MSE loss function that applies the weight $\omega$ to the MSE loss for elements where the mask is 1 (above the threshold). The MSE loss remains unchanged where the mask is 0, enabling accurate prediction of the temperature distribution in the region of interest with elevated temperatures exceeding 50$^\circ C$. The weighted MSE loss function is defined as
\begin{linenomath}
\begin{equation}
    \mathcal{L}_1(x,\widetilde{x})=\frac{1}{n}\sum_{i=1}^{n}(x_i-\widetilde{x}_i)^2\left(\omega m_i+(1-m_i)\right).
\end{equation}
\end{linenomath}
The Dice$_{(>50)}$ loss function is defined as
\begin{linenomath}\begin{equation}
    \text{Dice}_{(>50)}(x,\widetilde{x}) = 1-\frac{2\sum_{i=1}^{n}{x_{i~(>50)}\widetilde{x}_{i~(>50)}}}{\sum_{i=1}^{n}{x_{i~(>50)}^2}+\sum_{i=1}^{n}{\widetilde{x}_{i~(>50)}^2}}.
\end{equation}\end{linenomath}
In particular, the Dice$_{(>50)}$ loss function relies on the temperature (50$^\circ C$) at which the cell death pattern transitions to a predominance of necrosis, as discussed in \cite{Zhang2018temp}.
The combined loss function is as shown below:
\begin{linenomath}\begin{equation}\label{eq:loss}
    \mathcal{L}(x,\widetilde{x})=\frac{\alpha}{n}\sum_{i=1}^{n}(x_i-\widetilde{x}_i)^2+\beta\mathcal{L}_1(x,\widetilde{x})+\gamma~\text{Dice}_{(>50)}(x,\widetilde{x}),
\end{equation}\end{linenomath}
where
$x$ and $\widetilde{x}$ are predictions and ground truth, respectively, and $\alpha$, $\beta$, and $\gamma$ are weights.

\section{Results}\label{sec:4}
In this section, we present the evaluation metrics employed, experimental validation, and evaluate the prediction accuracy and inference time of the proposed network models. To demonstrate the robustness of the developed network, validation was also conducted using unforeseen tumor image data obtained from a new patient.

\subsection{Evaluation metrics}\label{sec:4.1}
Root mean squared error (RMSE), mean absolute error (MAE), and Dice score are used to evaluate test outcomes for temperature distribution. Specifically, the Dice scores, represented as Dice$_{(>40)}$ and Dice$_{(>50)}$, correspond to the temperature maps exceeding 40$^\circ C$ and 50$^\circ C$, respectively. In addition, Dice score \cite{sorenson,dice}, Jaccard score \cite{jaccard}, and Hausdorff distance \cite{hausdorff} are used to evaluate test outcomes for ablation lesion. The Hausdorff distance $d_\text{H}(X,Y)$ between two sets $X$ and $Y$ is defined as follows:
\begin{linenomath}
\begin{equation}
    d_\text{H}(X,Y)=\max\left\{\max_{x\in X}{d(x,Y)}, \max_{y\in Y}{d(X,y)}\right\},
\end{equation}
\end{linenomath}
where
\begin{linenomath}
\begin{equation}
    d(a,B)=\min_{b\in B}{d(a,b)}.
\end{equation}
\end{linenomath}

\subsection{Experimental validation}\label{sec:4.2}
\begin{figure*}[htpb!]
    \centering
    \includegraphics[width=\textwidth]{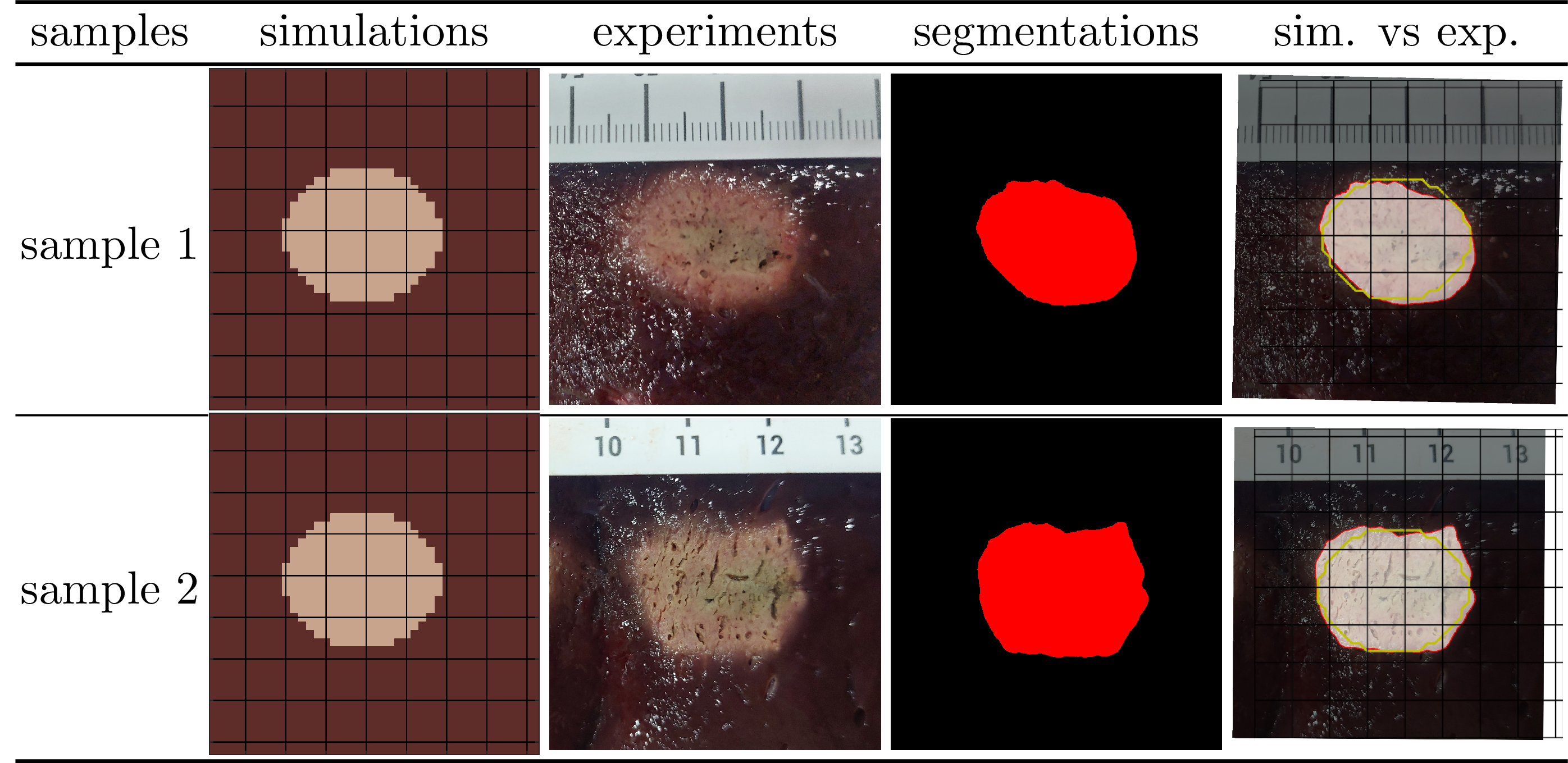}
    \caption{Comparison of simulated coagulative necrosis and RFA experiment results for two samples. The leftmost figures illustrate the numerical simulation outcomes. The second column represents experimental results. The third column shows segmented necrotic regions (red) obtained using SAM \cite{ma2023segment}. Meanwhile, the last column displays the segmented coagulative necrotic regions (red) with overlaid simulated results (yellow).}
    \label{fig:RFAexp}
\end{figure*}

In this section, we compare the outcomes of the RFA simulations with the RFA experimental results to validate the accuracy of our numerical RFA simulation. 
\figref{fig:RFAexp} presents a comparative analysis of two exemplary samples, selected from a total of seven,  demonstrating simulated coagulative necrosis alongside the results of RFA experiments. \tabref{tab:sim_exp_comparisons} provides length measurements, both horizontally and vertically, passing through the center of gravity of the necrosis region, along with its corresponding area. The center of gravity is determined using the \texttt{scipy.ndimage.center\_of\_mass} function \cite{scipy_cg}, and the horizontal and vertical lengths are obtained using ImageJ \cite{Schneider2012}. The results demonstrate an average accuracy within 2\% on the aspect of the area.

\begin{table}[htpb!]
\caption{Comparison of RFA Simulation outcomes and experimental segmentation data using diverse metrics. The term ``samples'' denotes the experimental sample numbers. ``horizontal'' and ``vertical'' refer to the lengths of lines, in mm. The ``area'', signifies the segmentation size in mm$^2$.}
    \centering
    \begin{tabular}{cccc}
    \toprule
       samples  & horizontal & vertical  & area \\
       \midrule
         1& 19.67 & 15.87&255.71\\ 
         2& 20.51 & 15.72&304.93\\ 
         3& 23.12 & 14.03&253.33\\ 
         4& 17.72 & 12.73&212.15\\ 
         5& 16.12 & 15.73&223.52\\ 
         6& 21.84 & 16.41&307.29\\ 
         7& 19.51 & 16.69&237.29\\ 
         \midrule
         avg$\pm$sd & 19.78$\pm$2.37 & 15.31$\pm$1.42&256.32$\pm$34.56\\
         \midrule
         simulation&20.00&16.00&261\\
         \midrule
         difference& 1.10\% & 4.31\%&1.79\%\\
         \bottomrule
    \end{tabular}

    \label{tab:sim_exp_comparisons}
\end{table}

\subsection{Evaluation of ablation lesion with foreseen/unforeseen test dataset}\label{sec:4.3}
We present an evaluation of the RFA lesion zone, demonstrating the performance of three different network architectures using both foreseen and unforeseen test datasets. \tabref{tab:damage_result} shows the prediction accuracy of damaged volume for each network model under the conditions of foreseen and unforeseen datasets. In the case of testing using a foreseen dataset, both U-Net and Attention U-Net architectures demonstrated higher predictive accuracy compared to EDCNN. Specifically, U-Net and Attention U-Net exhibited nearly identical accuracy with a Dice score of 96\%. On the other hand, the test using an unforeseen dataset showed slightly lower accuracy compared to the foreseen dataset tests. However, when utilizing the Attention U-Net architecture, it achieved the highest Dice accuracy of 93.6\% among the tested network models for the unforeseen dataset.

\begin{table}[htpb!]
    \centering
    \caption{Comparison of prediction accuracy for ablation lesion zone on foreseen/unforeseen test datasets. (num\_epochs = 100)}
    \begin{tabular}{llccc}
    \toprule
    test sets&networks&Dice&Jaccard&Hausdorff\\
    \midrule
    \multirow{3}{*}{foreseen}&EDCNN &0.9388&0.8871&1.1145 \\
    &U-Net &0.9624&0.9282&1.0202\\
    &Att. U-Net &0.9632&0.9298&1.0271\\
    \midrule
    \multirow{3}{*}{unforeseen}&EDCNN &0.9061&0.8338&1.3161 \\
    &U-Net &0.9185&0.8531&3.7220\\
    &Att. U-Net  &0.9364&0.8825&1.2255\\    
    \bottomrule
    \end{tabular}
    \label{tab:damage_result}
\end{table}

\begin{figure*}[htpb!]
    \centering
    \includegraphics[width=0.85\textwidth]{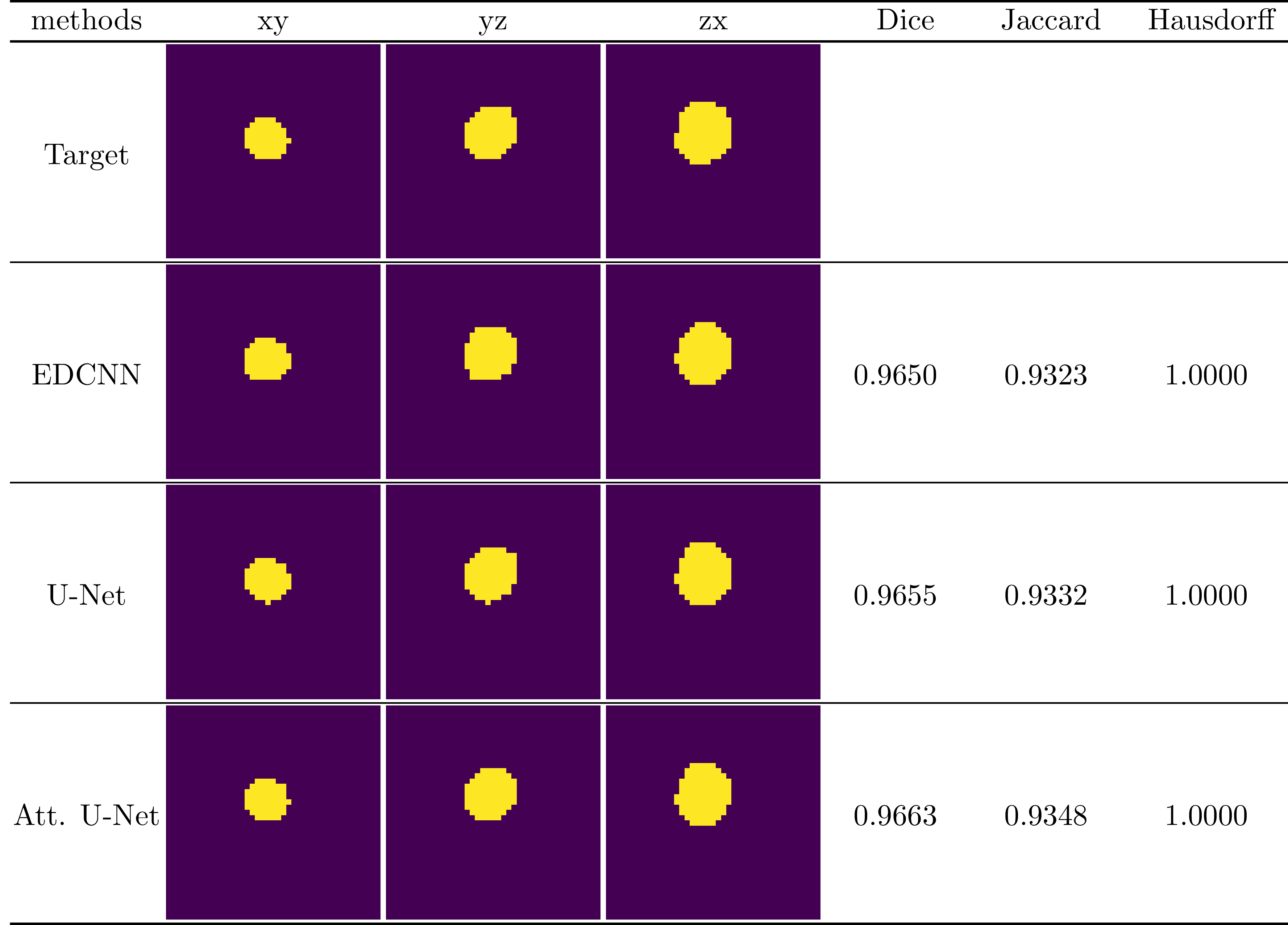}
    \caption{Comparison of ablation lesion predictions from various networks on an unforeseen test dataset with simulation results for a single sample. A positive trend in accuracy is observed in the order of EDCNN, U-Net, and Attention U-Net.}
    \label{fig:results_dmg}
\end{figure*}

\figref{fig:results_dmg} shows an exemplar graphical comparison of ablation lesion predictions with an unforeseen test dataset. The upward trend in accuracy is observed across EDCNN, U-Net, and Attention U-Net, with each subsequent model exhibiting improved performance over its predecessor. Among the three models, Attention U-Net exhibits the highest accuracy, indicating a progressive enhancement in the ability to analyze and interpret data correctly with each advanced model iteration.

\subsection{Evaluation of temperature distribution with foreseen/unforeseen test dataset}\label{sec:4.4}
Prior to assessing the performance of the presented network models in predicting temperature distribution, we conducted tests involving different weight distributions $(\alpha,\beta,\gamma)$ for the combined loss function defined in \eqref{eq:loss}. \tabref{tab:weights} indicates that $(\alpha, \beta, \gamma) = (0.7, 0.1, 0.2)$ yields the one of the best metrics. The selection of weights (0.7, 0.1, 0.2) is appropriate as minimizing the RMSE requires alpha to be the primary factor, dominating $\beta$ and $\gamma$. The small values of beta and gamma allow for slight refinements, particularly enhancing the accuracy in critical high-temperature regions. Therefore,  we adopted the determined weight distribution for the weighted loss function during the training process.

\begin{table}[htpb!]
\caption{Comparing the performance of loss functions using different weight combinations on the foreseen test dataset. The U-Net is employed for this evaluation. (num\_epochs=30)}
    \centering
    \begin{tabular}{ccccc}
    \toprule
    $(\alpha,~\beta,~\gamma)$&RMSE&MAE&Dice$_{(>40)}$&Dice$_{(>50)}$\\
    \midrule
    (1.0, 0.0, 0.0)&0.8201&0.5059&0.9766&0.9635\\ 
    (0.9, 0.1, 0.0)&0.9149&0.5827&0.9727&0.9561\\ 
    (0.9, 0.0, 0.1)&0.8938&0.5551&0.9755&0.9576\\ 
    (0.8, 0.2, 0.0)&0.8796&0.5419&0.9761&0.9607\\ 
    (0.8, 0.1, 0.1)&0.8278&0.4956&0.9779&0.9633\\ 
    (0.8, 0.0, 0.2)&1.0140&0.6689&0.9663&0.9469\\ 
    (0.7, 0.3, 0.0)&1.0001&0.5962&0.9754&0.9538\\ 
    (0.7, 0.2, 0.1)&0.8712&0.5463&0.9755&0.9579\\
    (0.7, 0.1, 0.2)&0.8153&0.5026&0.9766&0.9637\\ 
    (0.7, 0.0, 0.3)&0.9808&0.6550&0.9690&0.9462\\ 
    (0.6, 0.4, 0.0)&0.9285&0.5730&0.9739&0.9582\\ 
    (0.6, 0.3, 0.1)&0.9480&0.5852&0.9746&0.9525\\ 
    (0.6, 0.2, 0.2)&0.9036&0.5652&0.9733&0.9572\\ 
    (0.6, 0.1, 0.3)&1.0684&0.6888&0.9673&0.9488\\ 
    (0.6, 0.0, 0.4)&0.8569&0.5328&0.9755&0.9602\\ 
    \bottomrule
    \end{tabular}
    \label{tab:weights}
\end{table}

\tabref{tab:temperature_result} shows the prediction accuracy of temperature distribution for each network model under the conditions of foreseen and unforeseen datasets. In the case of a test using the foreseen datasets, it is observed that Attention U-Net achieves the highest accuracy among the three network architectures, with an RMSE of 0.4854. In the case of unforeseen results, there was a slight decrease in prediction accuracy compared to the foreseen test results. However, when utilizing the Attention U-Net architecture, a still notable accuracy of 0.6783 in RMSE was observed.

\begin{table}[htpb!]
    \centering
    \caption{Performance comparisons of temperature distribution on a foreseen/unforeseen test dataset.}
    {\footnotesize
    \begin{tabular}{p{1.4cm}p{1.4cm}p{0.7cm}p{0.7cm}p{0.7cm}p{0.7cm}}
    \toprule
    test sets&networks&RMSE&MAE&Dice$_{(>40)}$&Dice$_{(>50)}$\\
    \midrule
    \multirow{3}{*}{foreseen}&EDCNN&0.6653&0.4203&0.9779&0.9625\\      
    &U-Net&0.4853&0.2576&0.9887&0.9844\\
    &Att. U-Net&0.4854&0.2560&0.9892&0.9839\\       
    \midrule
    \multirow{3}{*}{unforeseen}&EDCNN&1.0502&0.7135&0.9564&0.9203\\    
    &U-Net&0.7340&0.4587&0.9793&0.9578\\
    &Att. U-Net&0.6783&0.3892&0.9845&0.9647\\      
    \bottomrule
    \end{tabular}}
    \label{tab:temperature_result}
\end{table}

\begin{figure*}[h!]
    \centering
    \includegraphics[width=0.85\textwidth]{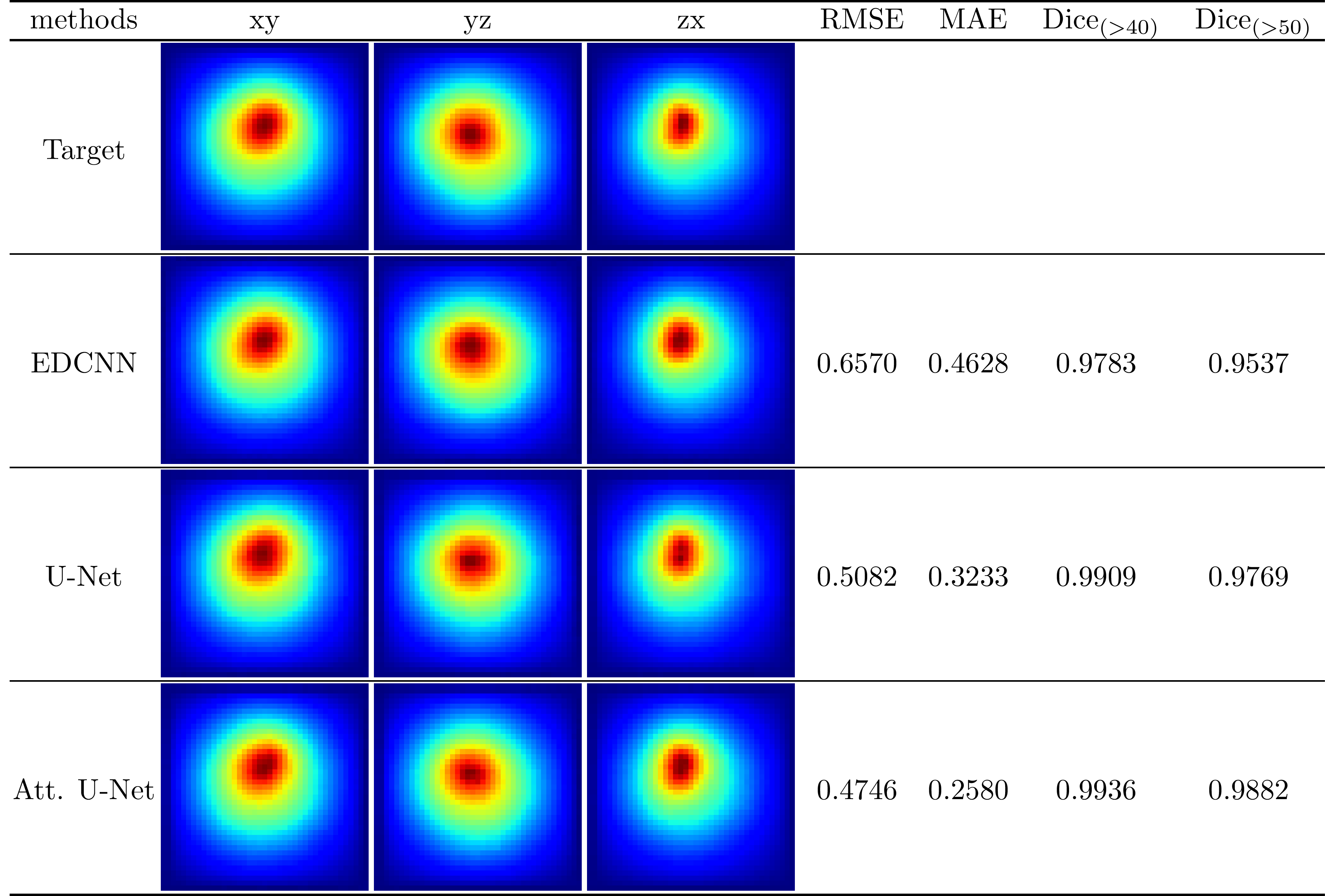}
    \caption{Comparison of temperature distribution predictions from various networks on an unforeseen test dataset with simulation results for a single sample.}
    \label{fig:results_temp}
\end{figure*}

\figref{fig:results_temp} shows an exemplar graphical comparison of temperature distribution predictions with an unforeseen test dataset. All three network models exhibit results closely resembling the target temperature distribution. Consistent with the results in \tabref{tab:temperature_result}, both U-Net and Attention U-Net demonstrate higher accuracy in all evaluation metrics compared to EDCNN. While the weighted Dice scores of U-Net and Attention U-Net are comparable, Attention U-Net shows a lower RMSE and MAE. This means that the temperature distribution greater than 50$^\circ C$ can be predicted with an accuracy of within 0.7\% by the use of Attention U-Net.

\begin{figure}[htpb!]
    \centering
    \includegraphics[width=0.45\textwidth]{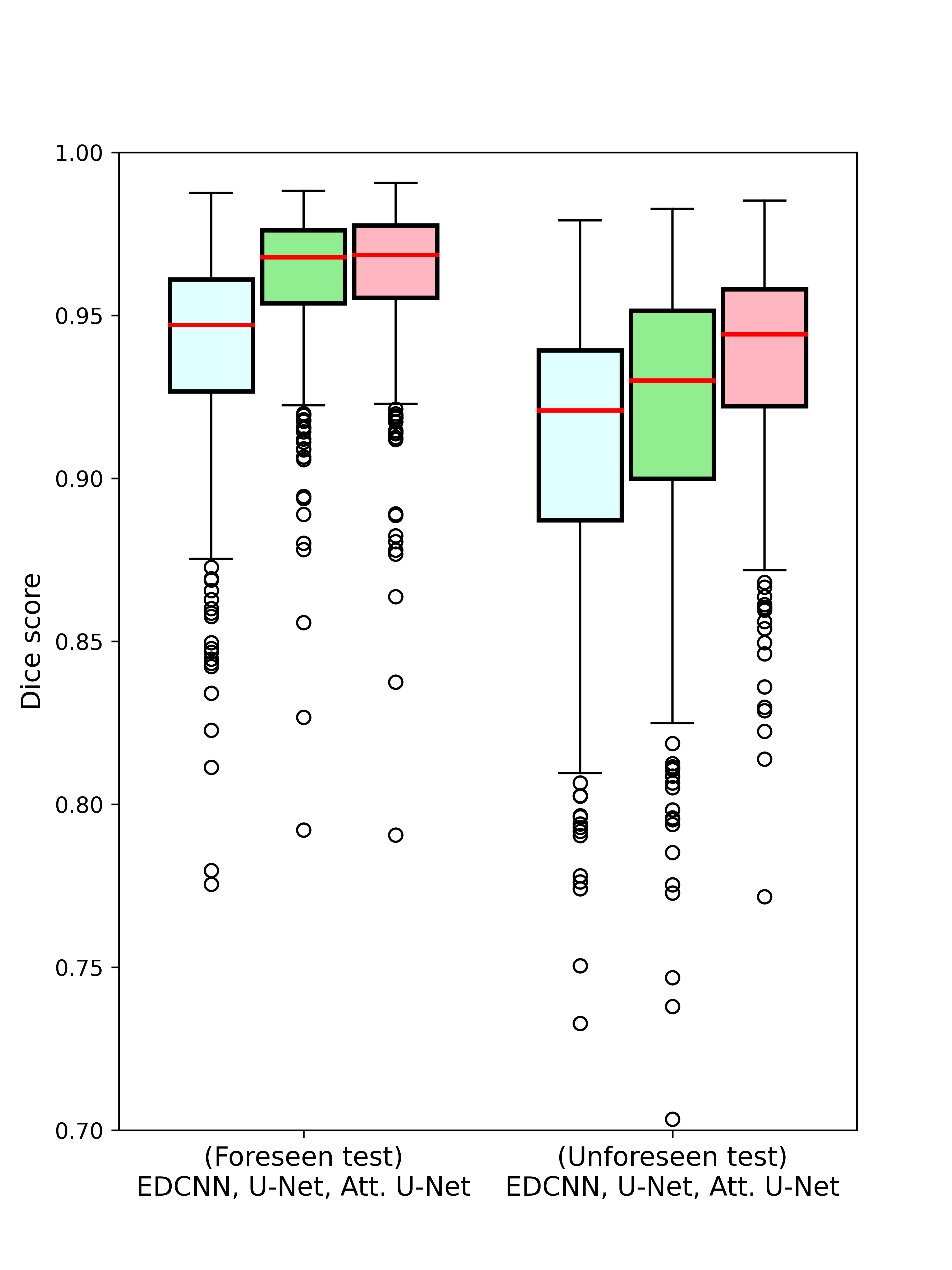}
    \caption{Dice score results for the ablation lesion zone with the methods of EDCNN, U-Net, and Attention U-Net. The three on the left were evaluated using a foreseen dataset, while the three on the right were assessed with an unforeseen dataset. The median value is denoted by the red horizontal line, while the interquartile range (IQR: from Q1 to Q3) is represented by the rectangle.}
    \label{fig:boxplot_Dmg}
\end{figure}

\begin{figure}[htpb!]
    \centering
    \includegraphics[width=0.45\textwidth]{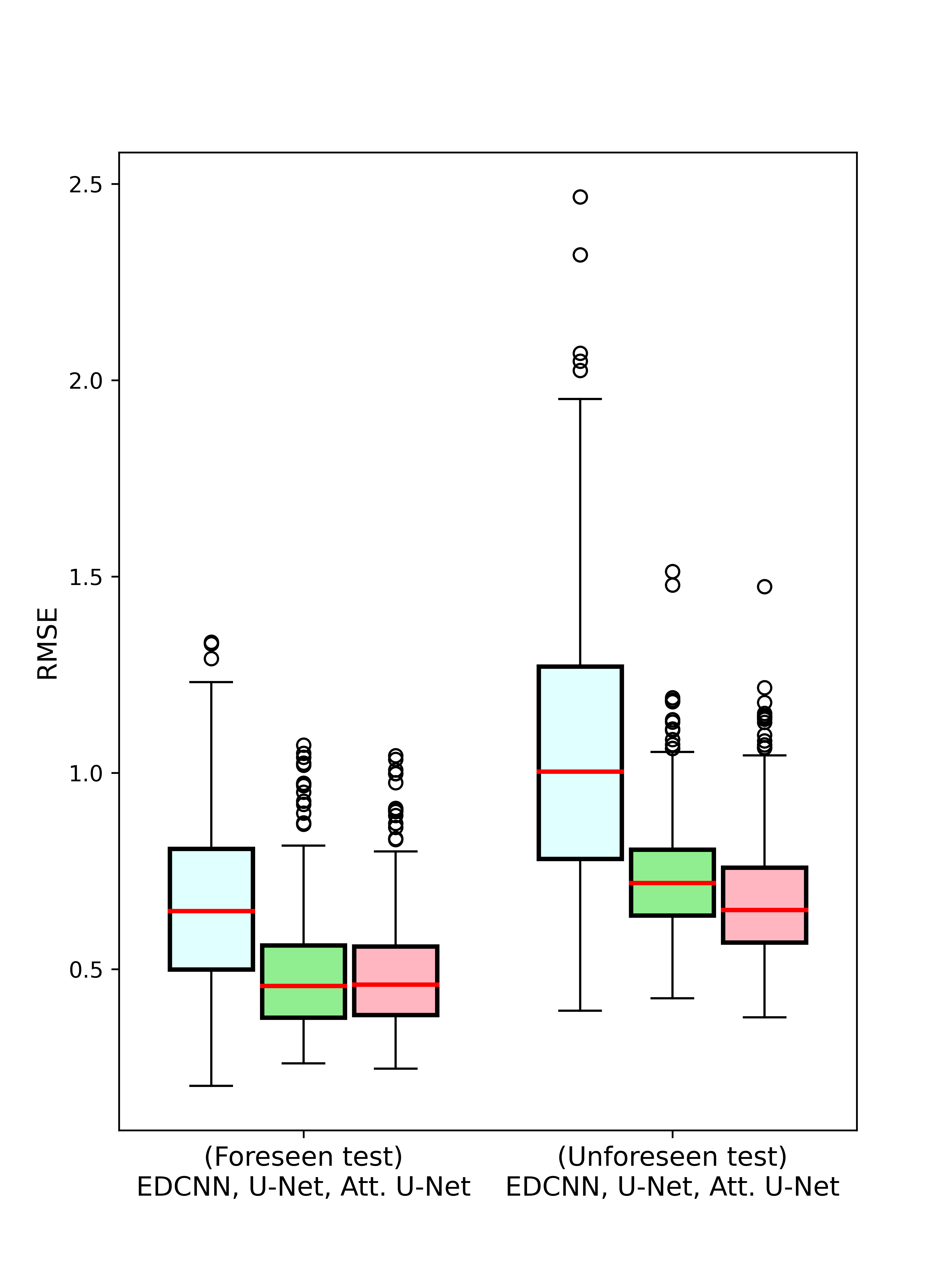}
    \caption{RMSE evaluations for the temperature distribution with the methods of EDCNN, U-Net, and Attention U-Net. The three on the left were evaluated using a foreseen dataset, while the three on the right were assessed with an unforeseen dataset. The median value is denoted by the red horizontal line, while the interquartile range (IQR: from Q1 to Q3) is represented by the rectangle.}
    \label{fig:boxplot_Temp}
\end{figure}

\figref{fig:boxplot_Dmg} and \figref{fig:boxplot_Temp} show the box plot of the ablation lesion volume and thermal distribution, respectively. In both cases, it is obvious that the order of prediction accuracy, from lowest to highest, is EDCNN, followed by U-Net, and then Attention U-Net.

\subsection{Evaluation of inference time}\label{sec:4.5}
In this section, we present a comparison of inference times across three different network architectures to assess their suitability for real-time inference. As shown in \tabref{tab:inference_time}, the inference times for EDCNN, U-Net, and Attention U-Net are approximately 1.49, 2.16, and 8.51 ms, respectively. This implies that the prediction of thermal effects (i.e., ablation lesion zone and temperature distribution) during RFA treatment can be completed in just under 10 ms with two inputs; the segmented breast tumor MR image and placement of ablation tip setup. The inference time was measured using parallel GPU computation in a desktop computer of AMD Ryzen 9 5900x 12-core processor 3.70 GHz and NVIDIA GeForce RTX 3090.

\begin{table}[htpb!]
    \centering
    \caption{Comparisons of inference times across different network architectures.}
    \begin{tabular}{lc}
    \toprule
    networks&inference time (ms)\\
    \midrule
    EDCNN& 1.4854      \\
    U-Net& 2.1557 \\ 
    Attention U-Net& 8.5140 \\ 
    \bottomrule
    \end{tabular}
    \label{tab:inference_time}
\end{table}

\section{Concluding remarks}\label{sec:5}
Digital twin, a representation of real-world objects, processes, and systems in a virtual space, stands at the forefront of the digital transformation era. In the healthcare domain, its application holds the potential to precisely monitor treatment processes and analyze patients' conditions, paving the way for personalized therapies and diagnostics. One crucial point in realizing this technology is the real-time simulation of physical phenomena occurring in the target object. Recent efforts have integrated mathematical analysis models with artificial intelligence (AI) techniques, resulting in the creation of AI models capable of real-time inference of physical phenomena. In this paper, we proposed physics-guided network models that can provide real-time feedback information on the thermal effects occurring during RFA treatment to users. The developed model achieved a 96\% accuracy in predicting the ablation lesion volume in terms of the Dice score and showed a temperature distribution difference of 0.4854 in terms of RMSE evaluation. Even in the evaluation using unforeseen data, the network demonstrated robust performance with a 93\% Dice score for ablation lesion prediction and a 0.6783 RMSE for temperature prediction. Furthermore, all network models exhibited real-time capabilities, inferring thermal effects within 10 ms (over 100 frames per second), using only standard home desktop computer specifications.

To assess the robustness of the networks for new patients, we conducted an accuracy evaluation using unforeseen data not involved in the training of the network model. As presented in \tabref{tab:damage_result} and \ref{tab:temperature_result}, the results indicated a minor decline in performance compared to results from data previously seen during training. Nevertheless, the networks still achieved high accuracy levels, highlighting the generality of the PhysRFANet and its capacity to adapt accurately to new breast tumor MR image data from new patients.

While the use of \textit{ex vivo} bovine liver tissue in the experiments provided validation of our models, it is crucial to acknowledge deficiency in replicating the dynamic physiological conditions inherent in living human patients \cite{KIM2011526}. The differences between bovine and human organs such as spatial relationships with other heat-sensitive organs, tissue cooling by the neighboring blood vessels \cite{Kunzli2011-zm}, blood flow dynamics \cite{Patterson1998}, immune response \cite{Mauda-Havakuk2022}, and other metabolic activities, may impact the generalizability of our results to human clinical scenarios. Despite this limitation, the use of \textit{ex vivo} bovine liver tissue offers a practical approach for preliminary investigations, laying the groundwork for further studies in more complex and clinically relevant settings.

This study did not account for RFA needle deflection or displacement \cite{dejong2018, perez2022} or deformation of the patient's breast tissue \cite{DANCHWIERZCHOWSKA2020} during RFA tip insertion. Neglecting these aspects could potentially impact the accuracy of the RFA procedure, as the tissue displacement resulting from changes in breast shape and needle deflection during tip insertion must be considered for precise treatment planning. Addressing these factors represents a promising avenue for future research endeavors.

To effectively utilize the developed models in clinical practice, it needs to be integrated with a technique for precisely placing RFA electrodes. For example, fusion imaging technology, specifically the integration of CT/MR-US fusion imaging, has significantly improved the accuracy of guiding RFA treatment \cite{McWilliams2010-am, Makino2016, WOOD2010S257, KRUCKER2011515}. The integrated approach of the PhysRFANet and fusion imaging technology can lead to a more targeted and effective application of RFA, thereby minimizing procedural errors and optimizing therapeutic outcomes.

In conclusion, this study introduces a novel approach that leverages physics-guided network models to provide real-time feedback on thermal effects during RFA treatment. Through MR images from 13 actual breast cancer patients, our approach has been verified to accurately infer patient-specific characteristics of the multi-biophysics phenomena occurring during RFA procedures. The proposed deep learning-based model holds great promise not only in advancing the standards of medical interventions and ensuring the safety and efficacy of RFA treatment but also in broadening the scope of RFA applications.

\section*{Acknowledgments}
The work was supported in part by the Korea Institute of Science and Technology (KIST) Institutional Program under Grant 2E31071, and in part by the National Research Foundation of Korea (NRF) grant funded by the Korean government (MSIT) (No. RS-2023-00220762).

\section*{Data and software availability}
The dataset, source code, and the description are available at
\href{https://github.com/iangilan/PhysRFANet}{https://github.com/iangilan/PhysRFANet}.

\section*{Declaration of competing interest}
The authors declare no competing interests.

\bibliographystyle{model1-num-names}
\bibliography{cas-refs}

\end{document}